\documentclass[conference]{IEEEtran}

\usepackage{graphicx,url}
\usepackage[caption=false,font=footnotesize]{subfig}
\usepackage{verbatim}
\usepackage{amsmath}
\usepackage{amssymb}
\usepackage{multirow}

\begin{document}

\title{Latency Versus Survivability in Geo-Distributed Data Center Design}

\author{
    \IEEEauthorblockN{Rodrigo S. Couto\IEEEauthorrefmark{1}, Stefano Secci\IEEEauthorrefmark{2}, Miguel Elias M. Campista\IEEEauthorrefmark{1}, and Lu\'is Henrique M. K. Costa\IEEEauthorrefmark{1}}
    \IEEEauthorblockA{\IEEEauthorrefmark{1}Universidade Federal do Rio de Janeiro - PEE/COPPE/GTA - DEL/POLI, Rio de Janeiro, RJ, Brazil\\
    Email:\{souza,miguel,luish\}@gta.ufrj.br}
    \IEEEauthorblockA{\IEEEauthorrefmark{2}Sorbonne Universit\'es, UPMC Univ Paris 06, UMR 7606, LIP6, F-75005, Paris, France\\
    Email:\{stefano.secci\}@upmc.fr}
}

\maketitle

\begin{abstract}
A hot topic in data center design is to envision geo-distributed architectures spanning a few sites across wide area networks, allowing more proximity to the end users and higher survivability, defined as the capacity of a system to operate after failures. 
As a shortcoming, this approach is subject to an increase of latency between servers, caused by their geographic distances. 
In this paper, we address the trade-off between latency and survivability in geo-distributed data centers, through the formulation of an optimization problem.
Simulations considering realistic scenarios show that the latency increase is significant only in the case of very strong survivability requirements, whereas it is negligible
for moderate survivability requirements. For instance, the worst-case latency is less than 4~ms when guaranteeing that 80\% of the servers are available after a failure, in a network where the latency could be up to 33~ms.
\footnote{This paper was published in the Proceedings of the IEEE GLOBECOM 2014 conference. The final version is available at http://dx.doi.org/10.1109/GLOCOM.2014.7036956. All rights are reserved to IEEE. \copyright 2014 IEEE.}
\end{abstract}

\section{Introduction}

Cloud computing can take advantage from server consolidation provided by virtualization, for management and logistic aspects. Virtualization also allows the distribution of Cloud servers across many sites, for the sake of fault-recovery and survivability. 
Indeed, the current trend in data center (DC) networking is to create a geographically-distributed architecture spanning a few sites across wide area networks. This allows to reduce the Cloud access latency to end users~\cite{develder2012optical}.
In addition, geo-distribution can guarantee high survivability against site failures or disconnection to clients.
Virtual machines (VMs) can in this way span various locations, and be instantiated as a function of various performance and resiliency goals, under an appropriated Cloud orchestration.
The main motivations to build up a geo-distributed DC are typically:
\begin{itemize}
 \item the achievable increase in DC survivability, hence availability and reliability;
 \item the reduced Cloud access delay thanks to the closer loop with some users;
 \item the possibility to scale up against capacity constraints (electricity, physical space, etc.).
\end{itemize}
We concentrate our attention on the survivability aspect.
The survivability of a Cloud fabric can be increased by distributing the DC over many sites, as much as possible, inside a given geographical region. The larger the geographical region, the lower the risk of 
multiple correlated failures such as backbone cable cuts, energy outages, or large-scale disasters~\cite{grover2004mesh}. 

Although DC distribution has positive side-effects, the design decision on the number and location of DC sites needs to take into consideration the increase of network latency between Cloud servers located in different sites, 
and the cost to interconnect sites in wide area networks (WANs). 
The latter aspect typically depends on various external factors, such as the traffic matrix inside the DC and CAPEX considerations. The former is of a more operational nature and becomes increasingly important in Cloud networking as 
even a few milliseconds of additional delay can be very important in the delivery of advanced services~\cite{develder2012optical}.
In this work, we concentrate on the trade-off between survivability and interconnection latency in the design of geo-distributed DCs.
Hence, we model the DC design as an optimization problem meeting both survivability and latency goals. 

In a general picture, survivable geo-distributed DC design recently started to be addressed in the literature, focusing on optical fiber capacity provisioning between DC sites~\cite{develder2012resilient,xiao2014Joint,habib2012design}. A common characteristic of these works is that they propose optimization problems to minimize the cost to provision network capacity and physical servers, leaving survivability as a constraint. 
Also, they assume that all services and demands are known at the time of DC design. 
The propagation delay caused by geo-distribution is only considered in~\cite{xiao2014Joint}, although it does not provide an explicit analysis of the trade-off between latency and survivability.
Our work adds to the state of the art in that we optimize both latency and survivability to assess their trade-off and answer to different survivability/latency requirements.
Hence, we isolate these two metrics by ignoring other factors such as physical costs (i.e., bandwidth utilization and cost to build a DC site).
Furthermore, our conclusions are independent of the services supported by the DC and of the traffic matrix generated by them.
We claim that the physical cost and the traffic matrix are undoubtedly important factors to consider in DC design. 
However, these factors should be ignored at a first approximation to better analyze the trade-off between latency and survivability.

Our simulation results show that in irregular mesh WANs a moderate level of survivability can be guaranteed without compromising the latency. Considering all of the analyzed networks, we find DC designs that guarantee that 80\% of all racks stay operational after any single failure, while increasing the maximum latency by only 3.6~ms when compared to the extreme case of a single DC site. On the other hand, very strong survivability requirements might incur in a high latency increase. 
For instance, by increasing the survivability guarantee from 94\% to 96\% of racks considering single failure cases, we observe that the latency may increase by 46\%.

This paper is organized as follows. Section~\ref{dcnmode} presents our modeling choices and design criteria, and Section~\ref{sec:problem} describes the related optimization problem.
Section~\ref{eval} reports our simulation results and Section~\ref{conc} concludes this work.

\section{Data-Center Network Model}
\label{dcnmode}

Our DC model is based on the following assumptions:

\begin{itemize}
\item the smallest DC unit is a rack, which consists of several servers connected by a top-of-rack (ToR) switch. The servers also have their external traffic aggregated by the ToR switch;
\item there are many DC site candidates in a given geographic region where we can install a given number of racks. We call the site \textit{active} if it has at least one installed rack;
\item Cloud users access DC services through gateways (Internet dependent or private) spread across the WAN;
\item only some DC sites have gateways for Cloud access traffic, as it happens in the current practice;
\item we account for link and node failures, where a node is a WAN switch/router or a DC site, and a link is the physical medium interconnecting a pair of nodes. Each link or node belongs to one or more Shared Risk Groups (SRGs), where an SRG is defined as a group of elements susceptible to a common failure~\cite{habib2013disaster}. For example, an SRG can be composed of DC sites attached to the same electrical substation. 
\end{itemize}

\begin{figure}
\centering
\includegraphics[width=0.48\textwidth]{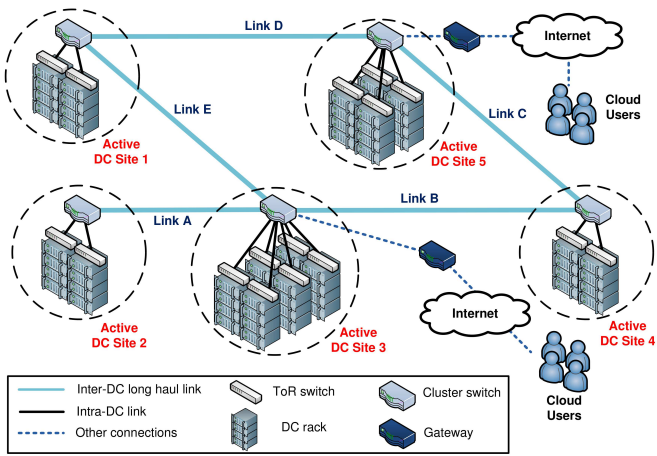}
\caption{Example of a geo-distributed DC composed of interconnected sites.}
\label{fig:scenario}
\end{figure}

Figure~\ref{fig:scenario} depicts an example of the reference scenario, where DC sites can host a different number of racks.
Depending on the WAN topology and on the density of gateways, different survivability and interconnection latency levels can be achieved, as described in the next paragraphs.

Our model captures the characteristics of a DC delivering IaaS (Infrastructure as a Service) as main service type. In the IaaS model, a DC provider allocates VMs to 
its clients, which in turn can have the possibility to remotely manage their IaaS to some extent (computing resources allocation, virtual linking, automated orchestration or manual migrations, etc.). 
A rack can host VMs from different clients, and each client shall have multiple VMs distributed across different racks and DC sites to improve survivability and flexibility. 

As the VMs of a given client are potentially geo-distributed, their availability can be increased by means of proactive or reactive migrations after failures.
The DC geo-distribution therefore meets IaaS availability requirements by allowing distribution of computing resources over different sites. 
However, since IaaS's VMs may intensively communicate with each other, the geo-distribution increases the distance
between racks and thus can impact the performance of the applications running on the VMs.
Obviously, a Cloud infrastructure where this tradeoff is well adjusted can provide better VM allocation to its clients, considering survivability and latency requirements. 
This is a constant concern in Cloud networking, especially for storage area network communications, where even an increase of 1 ms in RTT can represent a substantial performance degradation given the stringent I/O delay requirements~\cite{telikepalli2004storage}.
Next, we detail the two goals analyzed in this work. The main notations used in this work are provided in Table~\ref{tab:notations}.
\begin{table}
\caption{Notations}
\centering
\begin{center}
\begin{tabular}{|c||l|}
\hline $\mathcal{D}$ &Set of candidate DC sites\\
\hline $\mathcal{F}$ &Set of SRGs\\
\hline \multirow{2}{*}{$M_{fi}$} &Binary parameter indicating whether the SRG $f$ \\ 
&disconnects the site $i$ from the network\\
\hline $\Delta_{ij}$ &Propagation delay between the sites $i$ and $j$\\
\hline $u_i$    &Binary variable indicating if DC site $i$ is
active\\
\hline $R$ &Total number of racks to distribute among the DC sites\\
\hline $Z_i$ &Capacity (maximum number of racks supported) of site $i$\\
\hline $L_{max}$ &Maximum possible value of $l$ for a given topology\\
\hline $\beta$ &Parameter to weight $l$ versus $s$\\
\hline $s$			& Survivability \\
\hline $l$			& Interconnection latency \\
\hline $x_i$ 		& Number of racks in location $i$ \\
\hline
\end{tabular}
\label{tab:notations}
\end{center}
\end{table}

\subsection{Survivability goal}

To quantify the survivability of a geo-distributed DC, we use the concept of \lq\lq worst-case survival\rq\rq \ defined in~\cite{bodik2012surviving} as the fraction of DC services remaining operational after the worst-case failure of an SRG.
In the following, we thus refer to \lq\lq survivability\rq\rq \ as \textit{the smallest fraction of total racks available after the failure of any single SRG}.
In fact, by using the smallest fraction of racks, we consider the worst-case failure. 
We believe that this definition of survivability is very appropriate for geo-distributed DCs providing IaaS, since having less than 100\% of connected racks does not necessarily imply a degraded availability, i.e., a given service cannot be delivered. Formally, our survivability metric is computed as:
\begin{equation}
s = \min_{f \in \mathcal{F}} \left ( \frac{\sum_{k \in \mathcal{A}_f}r_k}{R} \right ),
\label{eq:survivability}
\end{equation}
where $\mathcal{F}$ is the set of all SRGs, 
$R$ is the total number of racks distributed among the DC sites, $\mathcal{A}_f$ is the set of accessible subnetworks after the failure of SRG $f$, and 
$r_k$ is the number of racks in the accessible subnetwork $k \in \mathcal{A}_f$. An accessible subnetwork is defined as a subnetwork that is isolated from the other subnetworks, but that has at least one gateway to the outside world. Recall that, after a failure, the network can be partitioned into different subnetworks. 
If a subnetwork does not have a gateway, its racks are not accessible from the outside world and thus cannot deliver DC services. 
For example, if in Figure~\ref{fig:scenario} the links B and D fail, the network is split in two accessible subnetworks: one composed of DC sites 1, 2, and 3; the other composed of DC sites 4 and 5.
Considering another failure scenario, where only link A fails, the network is also split in two subnetworks. One subnetwork is accessible and composed of DC sites 1, 3, 4, and 5. The other subnetwork, composed of DC site 2,
is not accessible since it does not have a path to a gateway.
Hence, $\sum_{k \in \mathcal{A}_f}r_k$ in Equation~\ref{eq:survivability} is the number of accessible racks after failure of SRG $f$.

According to this definition, the survivability metric assumes values in the interval $[0,1]$. It assumes the minimum value (zero) when all DC racks are in sites affected by a single SRG. The maximum value (one) occurs when the network has a certain level of redundancy and the DC is distributed in such a way that no single SRG can disconnect a rack.

\subsection{Interconnection latency goal}

We consider that the DC interconnection latency is mainly due to the inter-DC path's propagation delay.
That is, we consider that the network is well provisioned and that the latency due to congestion and retransmissions at intermediate nodes is negligible.
Under this assumption, we quantify the latency for DC interconnection as \textit{the maximum latency between pairs of active DC sites, considering all possible combinations}. 
Choosing the maximum value as reference metric is important to account for the fact that the VMs allocated to a given client may be spread over many sites. 
Thus, the maximum latency corresponds to the worst-case setting, where VM consolidation is conducted independent of site location or there is not enough capacity to perform a better VM allocation.
Formally, our latency metric is computed as:
\begin{equation}
l = \max_{i,j \in \mathcal{D}} ( \Delta_{ij} u_i u_j ),
\label{eq:latency}
\end{equation}

\noindent
where $\mathcal{D}$
is the set of DC sites, $\Delta_{ij}$ is the propagation delay between the sites $i$ and $j$, as defined above, and $u_i$ is a binary variable that is true if at least one rack is installed in site $i$. 
We consider that $\Delta_{ij} = \Delta_{ji}$, as paths using L1/L2 inter-DC WAN links are commonly set to be symmetric.
It is important to note that in the design problem described below, the interconnection latency is evaluated when there are no failures, to better analyze the trade-off between latency and survivability. Upon failure, alternative paths are chosen. If these paths have higher lengths, the latency increases.

\section{Problem Statement}
\label{sec:problem}

Our DC design problem has the twofold objective of maximizing the survivability while minimizing interconnection latency. It takes as parameters the latency between DC sites, the size of each DC site (i.e. the number of supported racks), SRG information, and the number of racks to allocate.
The output indicates how many racks to install in each site and, as a consequence, which DC sites to activate. The problem can be formulated as a Mixed Integer Linear Program (MILP) as follows, using the notations of Table~\ref{tab:notations}:

\begin{equation}
\textup{maximize}\ \  (1-\beta) s - \beta \frac{l}{L_{max}} 
\label{lp:objective}
\end{equation}

\begin{equation}
\textup{subject to} \ \ \ \sum_{i \in \mathcal{D}}M_{fi} x_i - sR \geq 0 \quad \forall f \in \mathcal{F}.
\label{lp:survivability}
\end{equation}
\begin{equation}
\begin{split}
l -\Delta_{ij}u_i -\Delta_{ij}u_j \geq -\Delta{ij} \\ \forall i,j \in \mathcal{D}, \ i < j.
\end{split}
\label{lp:latency}
\end{equation}
\begin{equation}
R u_i - x_i \geq 0 \quad \forall i \in \mathcal{D}.
\label{lp:binaryConversion1}
\end{equation}
\begin{equation}
u_i \leq x_i \quad \forall i \in \mathcal{D}.
\label{lp:binaryConversion2}
\end{equation}
\begin{equation}
\sum_{i \in \mathcal{D}} x_i = R.
\label{lp:totalRacks}
\end{equation}
\begin{equation}
x_i \leq Z_i \quad \forall i \in \mathcal{D}.
\label{lp:maxCapacity}
\end{equation}
\begin{equation}
s \geq 0, \quad l \geq 0, \quad x_i \geq 0 \ \forall \ i \in \mathcal{D}.
\label{lp:bounds}
\end{equation}
\begin{equation}
s \in \mathbb{R}; \quad l \in \mathbb{R}; \quad u_i \in \{0,1\}, \ \forall \ i \in \mathcal{D}; \quad x_i \in \mathbb{Z}, \ \forall \ i \in \mathcal{D}.
\label{lp:domain}
\end{equation}

The objective given by Equation~(\ref{lp:objective}) maximizes the survivability $s$, as defined in Equation~\ref{eq:survivability}; whereas it minimizes the latency $l$, as defined in Equation~\ref{eq:latency}.
The trade-off between latency and survivability is adjusted in Equation~(\ref{lp:objective}) by the scaling weight $0 \leq \beta \leq 1$. 
Also, $L_{max} = \max_{i,j \in \mathcal{D}} ( \Delta_{ij})$ is used to normalize $l$ to the interval $[0,1]$. Hence, both $l$ and $s$ assume values within the same interval.

Since Equations~\ref{eq:survivability}~and~\ref{eq:latency} are nonlinear, we linearize them as in Equations~(\ref{lp:survivability})~and~(\ref{lp:latency}), respectively. 
For survivability, applying Equation~(\ref{lp:survivability}) is equivalent to set $s$ to be less than or equal to the value defined by Equation~\ref{eq:survivability}.
As Equation~(\ref{lp:objective}) tries to increase the survivability, $s$ will have the highest possible value, assuming the same value as in Equation~\ref{eq:survivability}. 
Using the same reasoning, Equation~(\ref{lp:latency}) forces $l$ to assume the maximum latency between two active DC sites. 
To force Equation~(\ref{lp:latency}) to consider only active DC sites, we use the binary variables $u_i, i \in \mathcal{D}$.
Hence, if either $u_i$ or $u_j$ are zero, the constraint is not effective (e.g., if $u_i=0$ and $u_j=1$, the constraint is $l \geq 0$). 
The constraints defined by Equations~(\ref{lp:binaryConversion1})~and~(\ref{lp:binaryConversion2}) are used to set $u_i = 0$ if $x_i=0$, and  $u_i = 1$ if $x_i>0$.
Equation~(\ref{lp:totalRacks}) is applied to force the total number of racks to place ($R$), while Equation~(\ref{lp:maxCapacity}) limits the number of racks ($x_i$) allowed in each site $i$, respecting its capacity $Z_i$. 
Finally, Equations~(\ref{lp:bounds})~and~(\ref{lp:domain}) specify, respectively, the bounds and domain of each variable.

The latency parameters $\Delta_{ij}$ are computed over the shortest paths between the DC sites $i$ and $j$.
The binary parameters $M_{fi}$ are evaluated by removing from the network
all elements (nodes and links) of SRG $f$. 
Then, we analyze each DC site $i \in \mathcal{D}$ to check if it has a path to a gateway. Obviously, if a DC site is on the analyzed SRG, it is considered disconnected. 

\section{Evaluation}
\label{eval}
\begin{figure}
\centering
\includegraphics[width=0.48\textwidth]{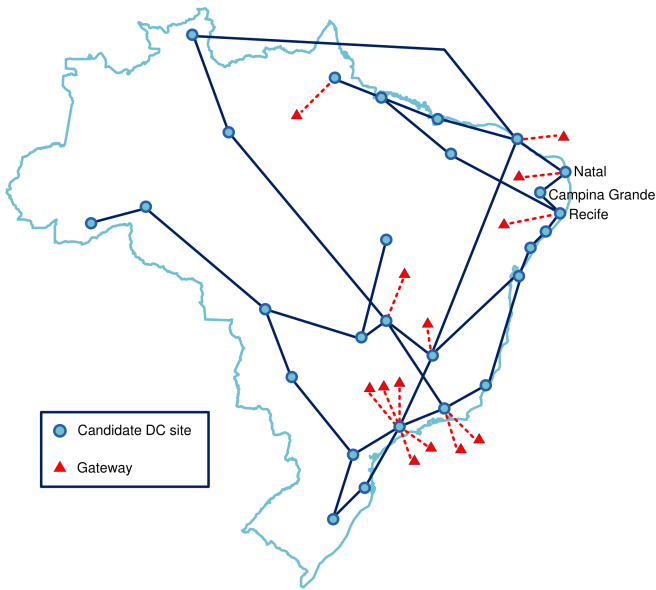}
\caption{RNP topology, with 27 sites and 33 links between them.}
\label{fig:redeIpe}
\end{figure}

\begin{figure}
\centering
\includegraphics[width=0.48\textwidth]{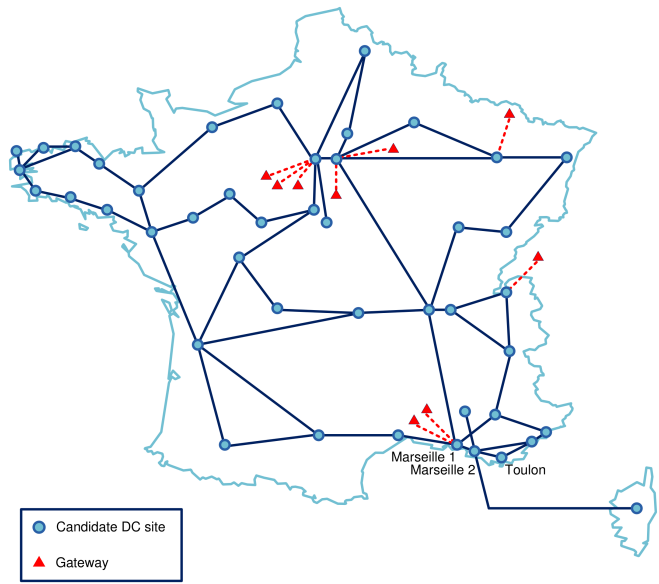}
\caption{RENATER topology, with 45 sites and 54 links between them.}
\label{fig:renater}
\end{figure}

\begin{figure}
\centering
\includegraphics[width=0.48\textwidth]{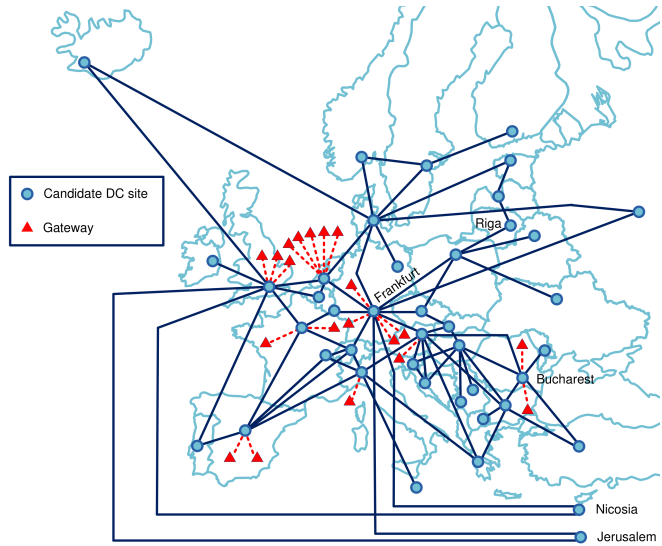}
\caption{GEANT topology, with 42 sites and 68 links between them.}
\label{fig:geant}
\end{figure}

We perform our analysis using real Research and Education Network (REN) topologies formed by many PoPs (Points of Presence).  
We thus consider that each PoP is a candidate DC site.
We adopt the WAN topologies of RNP (Figure~\ref{fig:redeIpe}) in Brazil, RENATER (Figure~\ref{fig:renater}) in France, and GEANT (Figure~\ref{fig:geant}) in Europe.
Each figure shows the DC sites and gateways of a given topology. For the sake of simplicity, we only specify in the figures the names of the sites that are mentioned along the text. 
Note that each topology covers a geographical area of different size. 
RNP and GEANT with respect to RENATER both cover a much larger area, with a surface more than 10 times larger than metropolitan France. 
However, RENATER has more nodes than RNP; whereas GEANT has a number of nodes close to RENATER. 


We use IBM ILOG CPLEX 12.5.1 as solver for the MILP problem. 
In addition, we adopt a single failure model, with one SRG for a single failure on either a DC site or a link.
The distance over a link is estimated as the length of a straight line between the centers of the two cities; the propagation delay is hence directly proportional to the distance, and we use a propagation speed of $2 \times 10^8$~m/s, which is 
a common assumption for optical fibers~\cite{ramaswami2009optical}. We use NetworkX~\cite{hagberg2008exploring} as graph analysis toolbox.
 
We set to 1024 the total number of racks $R$ to install. 
This number of racks was arbitrarily chosen, since the allocation depends on the relationship between $R$ and the capacity $Z_i$ of each site $i$ and not on any absolute values. 
In addition, we perform the evaluation for different $Z_i$ values (and considering, for simplicity, that all DC sites have the same capacity). 
Finally, to make the results independent of a specific $\beta$ value in Equation~(\ref{lp:objective}), the simulation is performed for the entire range of $\beta$ (specifically, from $0$ to $1$ in steps of $0.05$).

\begin{figure}
\centering
%generated by generate_rnp_betaVariation_combined.gnu
\subfloat[RNP, $L_{max}=32.37 \ \textup{ms}$.]
{\includegraphics[width=0.36\textwidth]{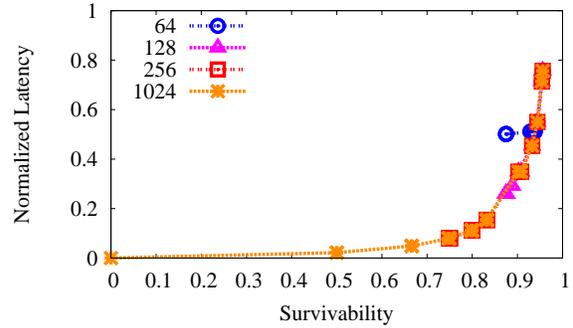}
\label{fig:latencyVsSurvivability_rnp}}\\
%generated by generate_renater_betaVariation_combined.gnu
\subfloat[RENATER, $L_{max}=7.12 \ \textup{ms}$.]
{\includegraphics[width=0.36\textwidth]{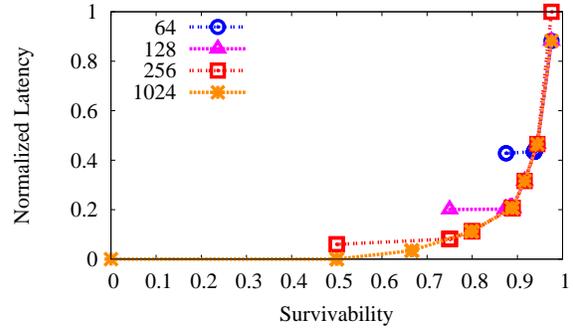}
\label{fig:latencyVsSurvivability_renater}}\\
%generated by generate_geant_betaVariation_combined.gnu
\subfloat[GEANT, $L_{max}=27.9 \ \textup{ms}$.]
{\includegraphics[width=0.36\textwidth]{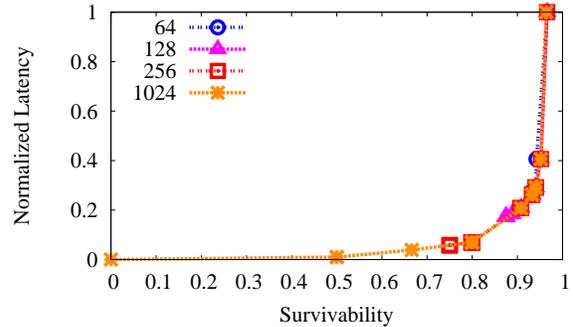}
\label{fig:latencyVsSurvivability_geant}}
\caption{Latency vs. survivability.}
\label{fig:latencyVsSurvivability}
\end{figure}

We plot in Figure~\ref{fig:latencyVsSurvivability} the normalized latency versus
the survivability for all networks, using the simulation data. The normalized latency is simply $\frac{l}{L_{max}}$, where $L_{max}$ is indicated in the captions corresponding to each topology. Recall that $L_{max}$ is the maximum latency $l$ computed over a shortest path between any two DC sites, active or not. 
Each curve in Figure~\ref{fig:latencyVsSurvivability} represents a different rack capacity $Z_i$ assigned to all DC sites (64, 128, 256, or 1024).
Note that, assigning a capacity of 1024, we assign full capacity to a single site since all racks can be put there.
Each data point in Figure~\ref{fig:latencyVsSurvivability}
is obtained by plotting the values of normalized latency and survivability achieved for the same $\beta$. 
For example, for RENATER in Figure~\ref{fig:latencyVsSurvivability_renater} with a capacity of 1024, we have a data point with a survivability of $0.89$ and a normalized latency of $1.47$, obtained using $\beta = 0.45$ in the simulation.
Note that the x-axis evolves on the opposite direction of $\beta$, since a larger $\beta$ increases the importance of the latency over the survivability. 

Results show a similar behavior for all topologies: for high survivability values, a small gain in survivability represents a high increase in latency. This happens because in all considered networks there are always a few nodes far away from the others. Therefore, when survivability requirements increases (i.e., $\beta$ decreases), these nodes are chosen because of lack of options. Thus, a slight improvement on survivability is achieved by inducing a severe increase in latency. 
As an example, for a full capacity (1024) setting and a survivability of $0.96$ in GEANT, the DC sites in Nicosia (Cyprus) and Jerusalem (Israel) are chosen, each one with 34 racks.
As shows Figure~\ref{fig:geant}, the path between these sites pass through the node in Frankfurt (Germany), having a total length of 5,581~km, which results in the maximum normalized latency of $1.00$ ($l = L_{max} = 27.9~ms$) as shown by Figure~\ref{fig:latencyVsSurvivability}. When the survivability decreases to $0.95$, the worst case for latency becomes the path between Riga (Latvia) and Bucharest (Romania). This path has a length of 2,267~km, which represents a normalized latency of $0.40$ ($l=11.33~ms$). 

\vspace{-4mm}
Conversely to the previous results we observe that, for lower survivability values, a significant increase in survivability produces a small increase in latency.
As an example, Figure~\ref{fig:latencyVsSurvivability_renater} shows that by varying the survivability from $0.5$ to $0.75$ with a rack capacity of 256, the normalized latency has a negligible increase from $0.06$ ($l=0.42$~ms) to $0.08$ ($l=0.58$~ms).
Hence, it shows that with very low latency values, the network can still achieve a moderate level of survivability.
Considering all networks, the maximum increase in latency when improving the survivability from 0 to 0.5 is the negligible value of 0.70~ms, which happens in RNP network. In this same network, a survivability of $0.80$ is achieved with only $0.11$ ($l=3.6$~ms) of normalized latency.
The low latency values achieved even with moderate levels of survivability are a consequence of a common characteristic among the considered networks: all of them have a high DC site density in a given region and they do not disconnect together. As an example, we can cite
the nodes in the northeast of Brazil (e.g., Natal, Campina Grande, Recife, etc.) and in the south of France (e.g., Toulon, the two nodes in Marseille, etc.).
We can thus spread DC racks in these regions without a significant latency increase.

The behavior highlighted above is valid for different site capacities, as seen in Figure~\ref{fig:latencyVsSurvivability}.
One difference is that, as we decrease the DC site capacity, the curves start from higher survivability values, 
since a lower capacity forces the DC to be more distributed into different geo-locations. For example, assigning a capacity of 64, we are imposing at least $\frac{1024}{64}=16$ active DC sites. However, this minimum number of active sites reduces the solution space an thus can lead to worse latency values. Hence, for some networks, the first data point (i.e., minimum survivability, achieved when $\beta=1$) has a higher latency as we decrease the capacity. After this point, all the data points lie on the full capacity curve for all networks.

In a nutshell, if the survivability requirement of a DC is not very high, it is easy to ensure moderate values of survivability without experiencing a significant latency increase in WAN mesh networks. 
However, moving the survivability to values close to 1 increases a lot the latency and may impact the performance of DC applications with tight latency requirements.
Note that these conclusions are valid for the three topologies, appearing to be independent of the geographical area covered by the WAN and of the number of nodes and links.
\begin{figure}
\centering
%generated by generate_rnp_betaVariation_DClocationsVsSurvivability.gnu
\subfloat[RNP]
{\includegraphics[width=0.36\textwidth]{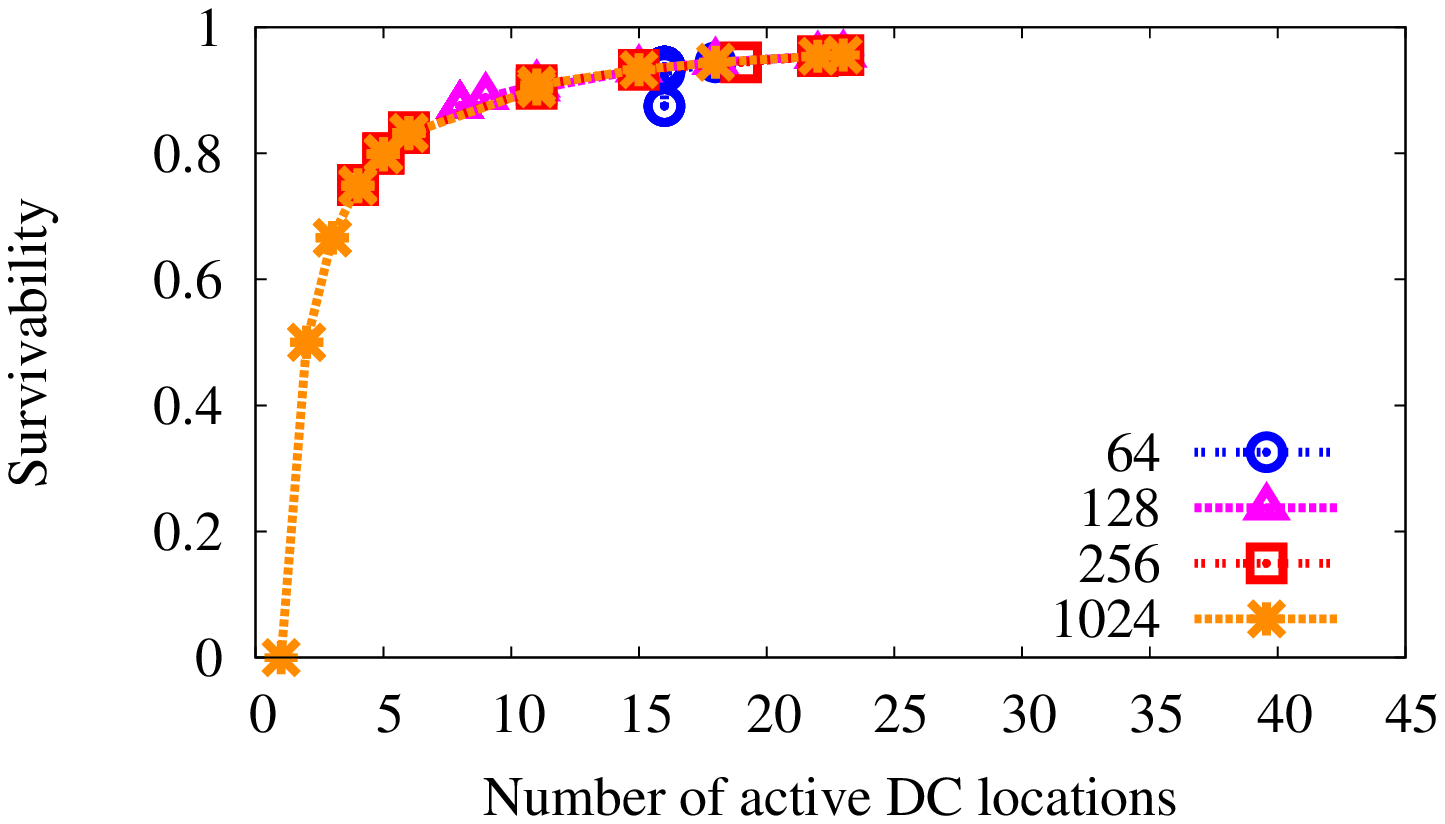}
\label{fig:dcLocationsVsSurvivability_rnp}}\\
%generated by generate_renater_betaVariation_DClocationsVsSurvivability.gnu
\subfloat[RENATER.]
{\includegraphics[width=0.36\textwidth]{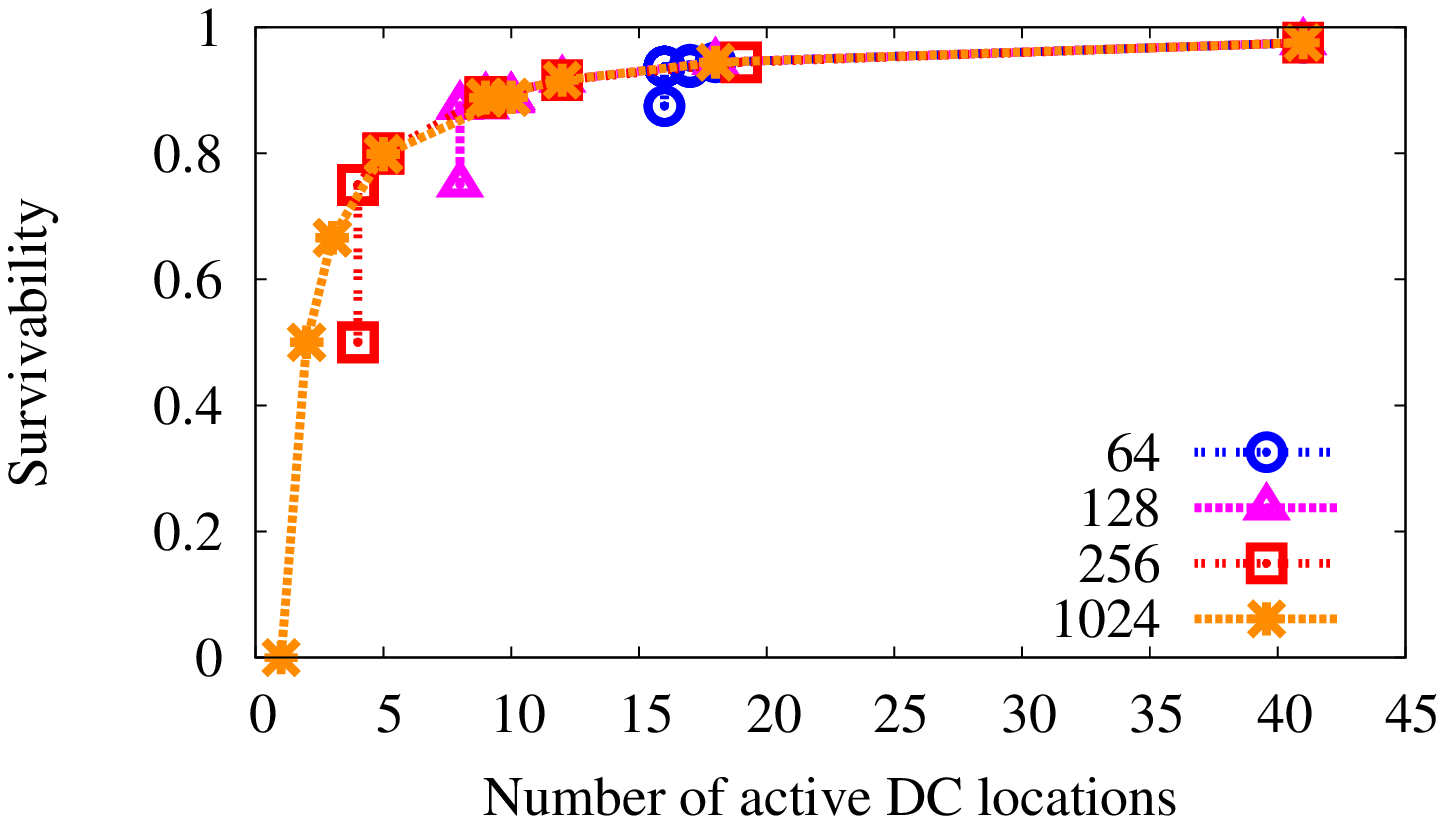}
\label{fig:dcLocationsVsSurvivability_renater}}\\
%generated by generate_geant_betaVariation_DClocationsVsSurvivability.gnu
\subfloat[GEANT.]
{\includegraphics[width=0.36\textwidth]{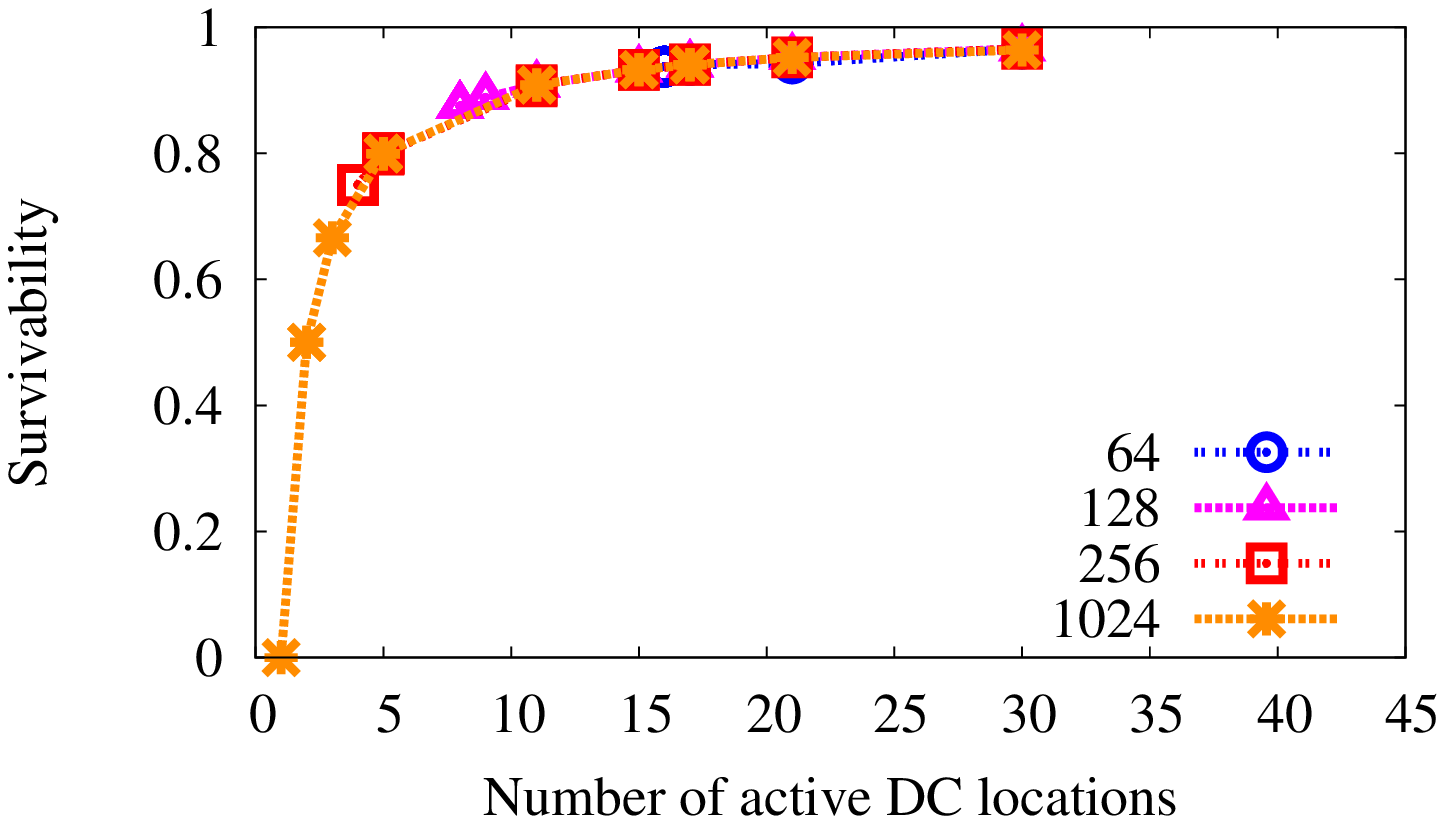}
\label{fig:dcLocationsVsSurvivability_geant}}
\caption{Survivability vs. number of active DC sites.}
\label{fig:dcLocationsVsSurvivability}
\end{figure}

Figure~\ref{fig:dcLocationsVsSurvivability} shows the survivability versus the number of active DC sites for the same experiment as above.
Although the optimization does not control directly this metric, the survivability growth is influenced by the DC distribution.
The DC distribution is influenced by the number of active DC sites. As more DC sites are used, the racks tend to span more SRGs
and thus the survivability tends to increase. The results show that the survivability grows logarithmically
with the number of active DC sites. This means that the increase in the number of active sites does not improve significantly the survivability when the DC becomes highly spread over different regions. 
Also, for a given capacity, the network can have different survivability levels for the same number of active DC sites. When observed, this behavior occurs only for the minimum possible number of active sites imposed by the capacity (e.g., $\frac{1024}{256}=4$ for a 256 capacity) achieved when $\beta=1$, where the survivability is not considered in the problem.

\section{Conclusions and Future Work}
\label{conc}

In this work we have analyzed the trade-off between latency and survivability in the design of geo-distributed DCs over WANs.
Simulations performed over diverse and realistic scenarios show that
when the DC survivability requirement is very high, a small improvement on the survivability produces a substantial increase in latency and hence a substantial decrease in IaaS performance. 
This is essentially due to the presence of very long paths for a few pair of nodes. At high survivability levels, our results show that an increase of 2\% (from 0.94 to 0.96) in the survivability level can increase the 
latency by 46\% (from 11.33~ms to 27.9~ms). On the other hand, when the DC survivability requirement stays moderate, the survivability can be considerably improved with a low latency increase. 
Considering all the WAN cases, the maximum latency increase is 0.7~ms when improving the survivability rate from 0 to 0.5. In addition, we observe in the considered WANs a maximum latency of 3.6~ms when the survivability is 0.8.

At present, the legacy situation is mostly characterized by monolithic DCs with a single or a very low number of sites, hence guaranteeing a very low survivability against site failures, as often experienced today by Internet users. 
Our study suggests that, considering a realistic scenario of DC design over wide area networks, increasing the geo-distributed DC survivability requirement to a moderate level only has little or no impact on IaaS delay performance.

As a future work, we plan to extend our study to include other objectives and constraints in the optimization problem, such as the network cost and the latency for end users.

\IEEEpeerreviewmaketitle

\section*{Acknowledgement}
This work was partially supported by FAPERJ, CNPq, CAPES research agencies, and the Systematic FUI 15 RAVIR (\url{http://www.ravir.io}) project.

\bibliographystyle{IEEEtran}
\bibliography{gc2014GeoDC_arxiv}

\end{document}